\documentclass[preprint,aps]{revtex4}

\usepackage{graphicx}% Include figure files
\usepackage{dcolumn}% Align table columns on decimal point
\usepackage{bm}% bold math
\usepackage{epsf}

\newcommand{\be}{\begin{eqnarray}}

\newcommand{\Eq}[1]{Eq.~(\ref{#1})}

\newcommand{\beq}{\begin{equation}}
\newcommand{\eeq}{\end{equation}}

\newcommand{\la}[1]{\label{#1}}
\newcommand{\bea}{\begin{eqnarray}}
\newcommand{\eea}{\end{eqnarray}}
\newcommand{\beqa}{\begin{eqnarray}}
\newcommand{\eeqa}{\end{eqnarray}}
\newcommand{\ba}{\begin{array}}
\newcommand{\ea}{\end{array}}

\newcommand{\half}{{\textstyle{\frac{1}{2}}}}

\def\appendix{\par
\setcounter{subsection}{0}
\setcounter{equation}{0}

\def\thesection{Appendix}
\def\theequation{\Alph{section}.\arabic{equation}}}

\begin{document}

\title{\bf Chiral symmetry and pentaquarks}
\vskip 0.5true cm

\author{\bf Dmitri Diakonov$^{1,2,3}$}
\vskip 0.5true cm

\affiliation{Thomas Jefferson National Accelerator Facility, Newport News, VA 23606, USA\\
NORDITA, Blegdamsvej 17, DK-2100 Copenhagen, Denmark\\
St.~Petersburg Nuclear Physics Institute, Gatchina, 188300, St.~Petersburg, Russia}
\vskip 0.5true cm

\date{December 18, 2004}
\vskip 0.5true cm

\begin{abstract}
Spontaneous chiral symmetry breaking, mesons and baryons are illustrated in the
language of the Dirac theory. Various forces acting between quarks inside baryons
are discussed. I explain why the naive quark models typically overestimate 
pentaquark masses by some $500\,{\rm MeV}$ and why in the fully relativistic approach 
to baryons pentaquarks turn out to be light. I discuss briefly why it can be easier to produce
pentaquarks at low than at high energies~\cite{F1}.
\end{abstract}

\maketitle

\section{On confinement}

Confinement of color may be realized in a way that is more subtle than some people think.
An example of a subtle confinement is provided by the exactly solvable Quantum
Electrodynamics in $1\!+\!1$ dimensions, also known as the Schwinger model. In the ``pure
glue'' variant of the model, {\it i.e.} with light ``quarks'' switched off, there is a
trivial linear confining potential between static external charges, since the Coulomb potential 
is linear in one dimension. However, as one switches in massless or nearly massless
``quarks'', the would-be linear confining potential of the imaginary pure-glue world is completely
screened: it is energetically more favorable to produce ``mesons'' than to pump an infinitely
rising energy into the ever-expanding string between the sources. Nevertheless, ``quarks''
are not observable in the Schwinger model: they are confined despite the absence of
gluonic strings or flux tubes between them. Only ``mesons'' are observable, built of
an indefinite number of quark-antiquark ($Q\bar Q$) pairs~\cite{Schwinger}. 

Turning to Quantum Chromodynamics in $3\!+\!1$ dimensions, there may be certain doubts
whether there actually exists a linear rising potential between static quarks in the pure
glue version of the theory (the systematic errors  for that potential measured in lattice 
simulations may be underestimated, especially for large separations where it is most 
interesting~\cite{lin-pot}), however in the real world with light $u,d,s$ quarks 
color strings or flux tubes between quarks undoubtedly do not exist. It is reassuring
that the screening of the rising potential has just started to be revealed in 
lattice simulations with light quarks~\cite{string-break}. Unfortunately, so far the
string breaking has been observed either at non-zero temperatures, or in $2\!+\!1$ 
dimensions, or on very coarse lattices: such computations are very time-expensive. 
It implies that all lattice simulations for the ``real'' QCD are at present running with 
inherent strings between quarks, which do not exist in nature! It means that either 
wrong physics is cured in the process of the extrapolation of the present-day
lattice results to small quark masses, or that the artifact strings are not too relevant
for most of the observables: for example, they may be effectively broken without
notice.   
\vskip -0.7true cm

\section{Spontaneous Chiral Symmetry Breaking (SCSB)}
 
Besides confinement, the other crucial aspect of QCD is the spontaneous breaking of
the chiral symmetry: as the result the nearly massless ``bare'' or ``current''
$u,d,s$ quarks obtain a dynamical, momentum-dependent mass $M(p)$ with $M(0)\approx
350\,{\rm MeV}$ for the $u,d$ quarks and $\approx 470\,{\rm MeV}$ for the $s$ quark. 
The microscopic origin of how light quarks become heavy, including the above
numbers, can be understood as due to instantons~\cite{DP-SCSB,inst-at-work} -- large 
fluctuations of the gluon field in the vacuum, needed to make the $\eta'(958)$
meson heavy~\cite{tHooft}. Instantons are specific fluctuations of the gluon field
that are capable of capturing light quarks. Quantum-mechanically, quarks can
hop from one instanton to another each time flipping the helicity. When it
happens many times quarks obtain the dynamical mass $M(p)$. This mass goes to zero 
at large quark virtuality since quarks with very high momenta are not affected by any 
background, even if it is a strong gluon field as in the case of instantons, see Fig.~1.
%%%%%%%%%%%%%%
%% FIGURE 1 %%
%%%%%%%%%%%%%%
\begin{figure}[b]
\centerline{\epsfxsize=5cm\epsfbox{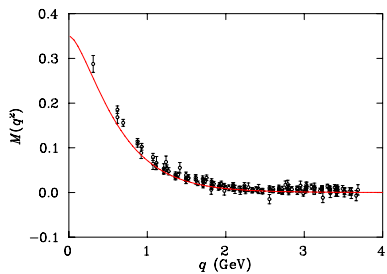}} 
\caption{Dynamical quark mass $M(p)$ from a lattice simulation~\cite{Mlat}. 
Solid curve: obtained from instantons two decades before lattice 
measurements~\cite{DP-SCSB}.}
\end{figure}
Instantons may not be the only and the whole truth but the mechanism of
the SCSB as due to the delocalization of the zero quark modes in the
vacuum~\cite{DP-SCSB} is probably here to stay.     

When chiral symmetry is spontaneously broken, the eight pseudoscalar mesons $\pi,K,\eta$
become light (quasi) Goldstone bosons. In the chiral limit ({\it i.e.} when the
bare quark masses $m_u\approx 4,\,m_d\approx 7,\,m_s\approx 150\,{\rm MeV}$ are set to
zero) the pseudoscalar mesons are exactly massless as they correspond to going along
the ``Mexican hat'' valley, which costs zero energy. For the future discussion of pentaquarks
it will be useful to understand chiral symmetry breaking in the language of the Dirac
sea of quarks, see Figs.~2,3~\cite{F2}.   
%%%%%%%%%%%%%%%%
%% FIGURE 2,3 %%
%%%%%%%%%%%%%%%%
\begin{figure}[htb]
\vskip -0.2true cm

\begin{minipage}[t]{.45\textwidth}
\vspace{-4.4cm}
\includegraphics[width=0.65\textwidth]{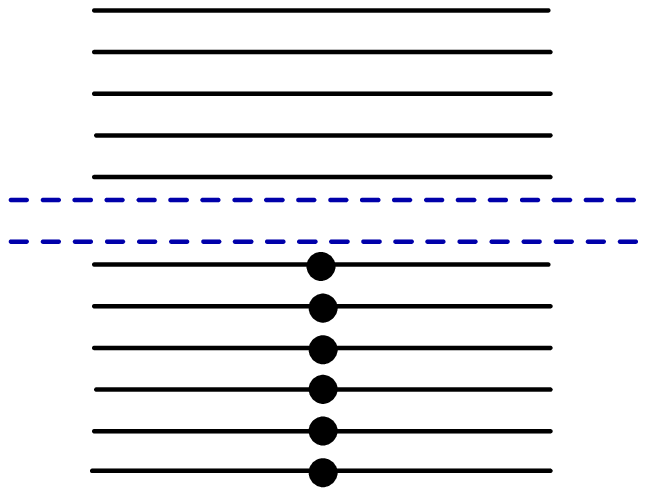}
\vskip 0.7true cm

\caption{Dirac spectrum of quarks {\it before} spontaneous chiral symmetry breaking.
Since quarks are massless or nearly massless, there is no gap between the positive-
and negative-energy Dirac continua.}
\label{Fig:2}
\end{minipage}
\hskip 1true cm %\hfil
\begin{minipage}[t]{.45\textwidth}
\includegraphics[width=\textwidth]{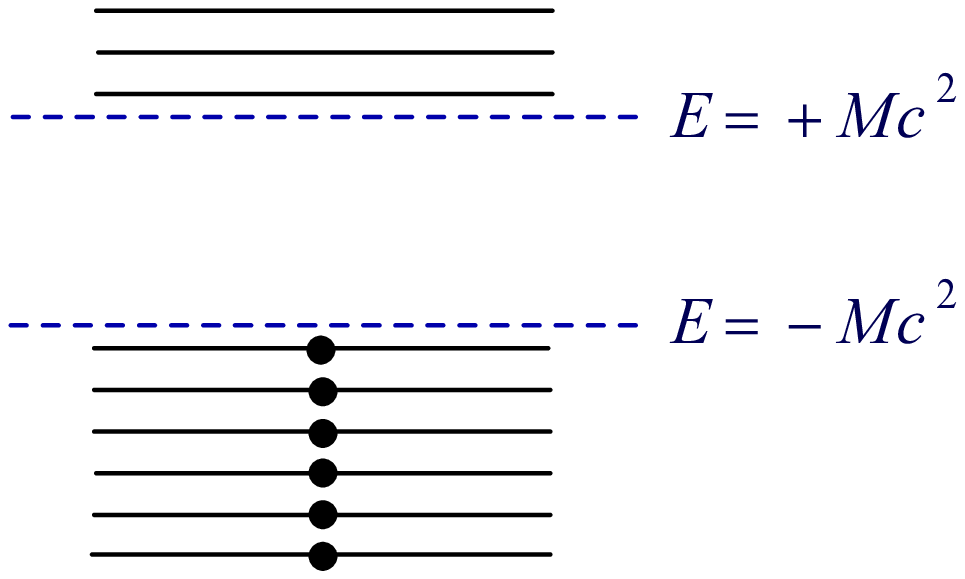}
\caption{Spontaneous chiral symmetry breaking makes a mass gap of $2Mc^2$ in the Dirac 
spectrum. The vacuum state (no particles) corresponds to filling in all negative-energy 
levels.}
\label{Fig:3}
\end{minipage}
\end{figure}

The appearance of the dynamical mass $M(p)$ is instrumental in understanding the world
of hadrons made of $u,d,s$ quarks. Indeed, the normal lowest lying vector mesons have 
approximately twice this mass while the ground-state baryons have the mass
of approximately thrice $M$. It does not mean that they are weakly bound: as usual in quantum
mechanics, the gain in the potential energy of a bound system is to a big extent compensated
by the loss in the kinetic energy of the localized quarks, as a consequence of the 
uncertainty principle. Therefore, one should expect the size of light hadrons to be 
on the scale of $1/M\approx 0.7\,{\rm fm}$, which indeed they are. At the same time 
the size of the constituent quarks is roughly given by the slope of $M(p)$ in Fig.~1, 
corresponding to about $\frac{1}{3}\,{\rm fm}$. Therefore, constituent quarks in hadrons 
are rather small in size and hence generally well separated, which is a highly non-trivial 
fact. It explains why the constituent quark idea has been a useful guideline for 40 years.

\section{Mesons}

In the language of the Dirac spectrum for quarks, vector, axial and tensor mesons  
are the particle-hole excitations of the vacuum, see Fig.~4. In the Dirac theory, 
a hole in the negative-energy continuum is the absence of a quark with negative energy, 
or the presence of an antiquark with positive energy. To create such an excitation, 
one has to knock out a quark from the sea and place it in the upper continuum: 
that costs minimum $2M$ in a non-interacting case, and gives the scale of the vector 
(as well as axial and tensor) meson masses in the interacting case as well. 

%%%%%%%%%%%%%%%%
%% FIGURE 4,5 %%
%%%%%%%%%%%%%%%%
\begin{figure}[htb]
\begin{minipage}[t]{.45\textwidth}
\includegraphics[width=1.0\textwidth]{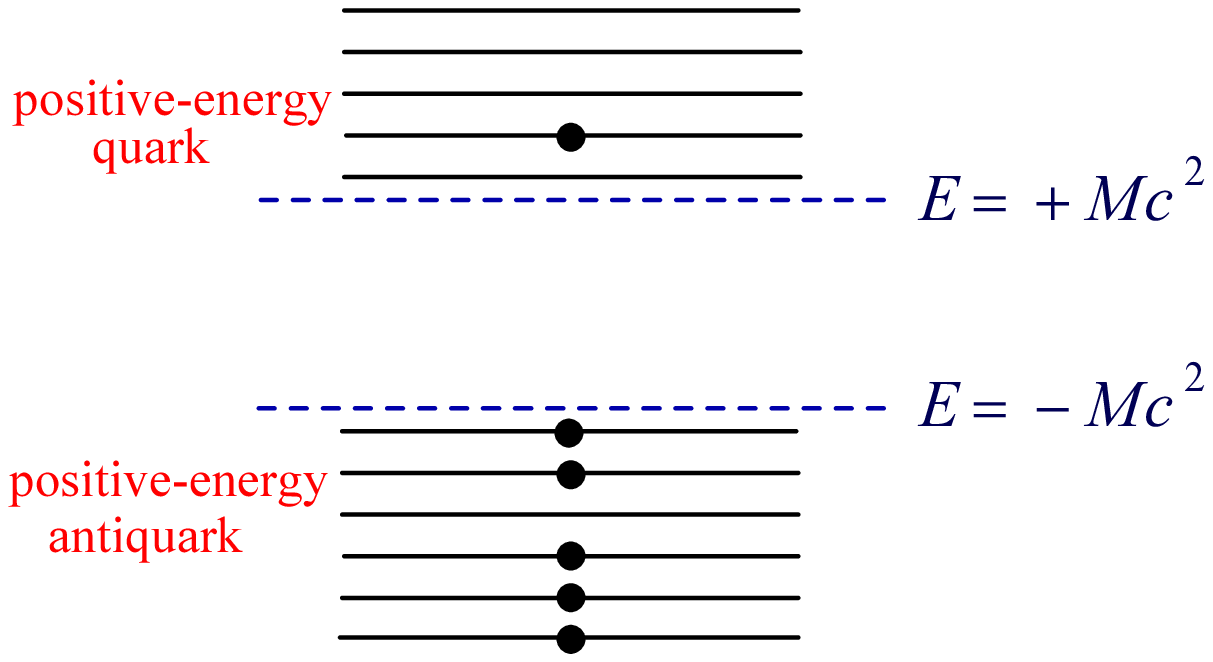}
\caption{Vector mesons are particle-hole excitations of the vacuum. 
They are made of a quark with positive energy and an antiquark with positive energy, 
hence their mass is roughly $2M$.}
\label{Fig:4}
\end{minipage}
\hfil
\begin{minipage}[t]{.45\textwidth}
\includegraphics[width=1.0\textwidth]{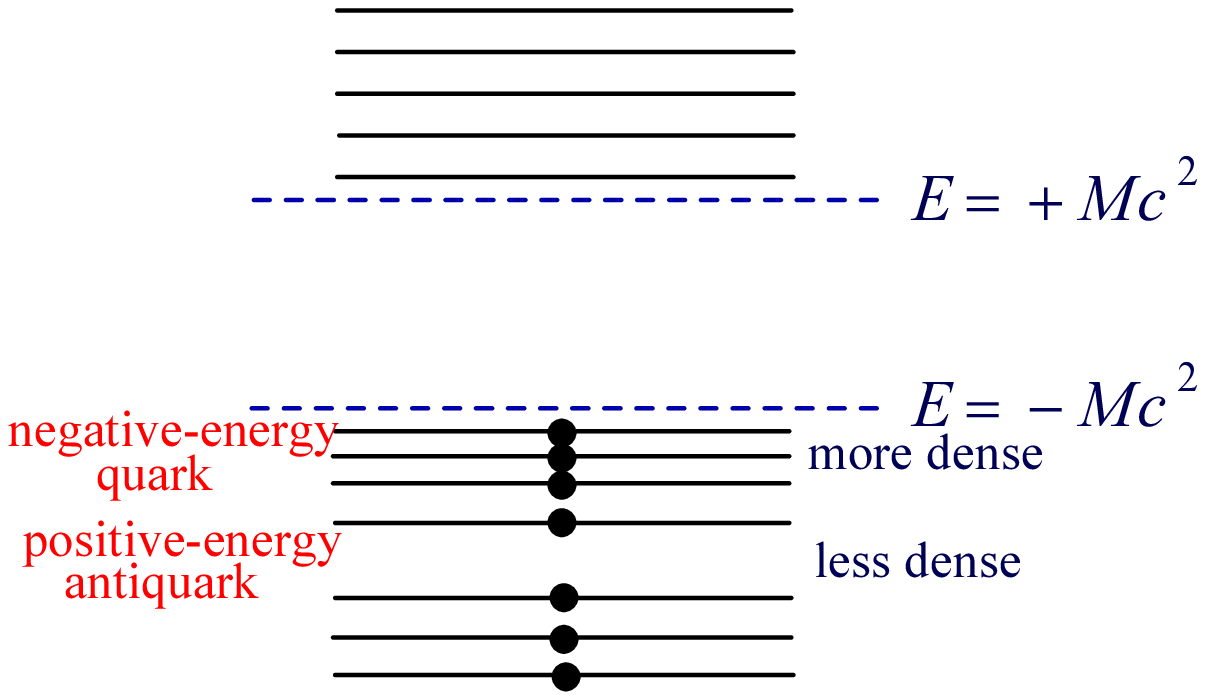}
\caption{Pseudoscalar mesons are {\it not} part\-icle -hole excitations but a collective
re-arran\-ge\-ment of the vacuum. They are made of an antiquark with positive
energy and a quark with {\it negative} energy, hence their mass is roughly zero.}
\label{Fig:5}
\end{minipage}
\end{figure}

For pions, this arithmetic fails: their mass is zero by virtue of the Goldstone 
theorem. One can say that in pions twice the constituent quark mass is completely eaten up
by a strong interaction (which is correct) but there is a more neat way to understand it.
 
Pseudoscalar mesons are totally different in nature from, say, the vector mesons. 
They are Goldstone bosons associated with symmetry breaking. Adding a pion to the vacuum 
is a chiral rotation which costs zero energy: it is the same vacuum state. Pseudoscalar 
mesons are described by the same filled Dirac sea with negative energies as the vacuum 
itself. They are not particle-hole excitations. If the Goldstone boson carries some energy, 
it corresponds to a slightly distorted spectral density of the Dirac sea (Fig.~5). 
The region of the Dirac sea where the level density is lower than in the vacuum, 
is a hole and corresponds to an antiquark with positive energy. The region with 
higher density than in the vacuum corresponds to an extra quark with a negative energy, 
since there are now ``more quarks'' in the negative-energy Dirac sea. Therefore, the 
pseudoscalar mesons are ``made of'' a {\em positive-energy antiquark} and a 
{\em {\bf negative}-energy quark}. Their mass is hence $(M - M) = 0$. This explains why 
the mass is zero in the chiral limit, or close to zero if one recalls the small $u,d,s$ 
bare masses which break explicitly chiral symmetry from the start.  

The most interesting mesons are the scalar ones: they are chiral partners of the pseudoscalar
mesons and their quark organization depends much on the concrete mechanism by which 
chiral symmetry is broken, in particular on the stiffness of the ``Mexican hat''. In the
instanton model of the QCD vacuum, the $Q\bar Q$ interaction in scalar mesons is much 
stronger than in vector, axial and tensor mesons. One can then expect the intermediate 
status of the scalar mesons, between Figs.~4 and 5. In addition, two pseudoscalar 
excitations from Fig.~5 may resonate forming a scalar resonance. Therefore, scalar mesons 
can be a mix of a positive-energy quark bound with a positive-energy antiquark, 
and two positive-energy antiquarks bound with two {\em negative}-energy quarks (and {\it vice versa}). 
Which component prevails is very difficult to predict without a detailed dynamical theory 
but the data seem to indicate~\cite{Achasov} that the lowest nonet 
($\sigma(600),\kappa(800),a_0(980),f_0(980)$) is predominantly a four-quark state 
(with two negative-energy quarks which make them unusually light!) whereas the second 
nonet ($f_0(1370),K^*_0(1430),a_0(1450),f_0(1530)$) are predominantly ``normal'' particle-hole 
mesons, although its singlet member can be already mixed with the gluonium. A recent
study of the lightest scalar nonet in the instanton liquid model~\cite{Schaefer} reveals
the above features.  

\section{Baryons}

Without spontaneous chiral symmetry breaking, the nucleon would be either nearly massless
or degenerate with its chiral partner, $N(1535,\half^-)$. Both alternatives are many hundreds of 
MeV away from reality, which serves as one of the most spectacular experimental indications 
that chiral symmetry is spontaneously broken. It also serves as a warning that if we disregard 
the effects of the SCSB we shall get nowhere in understanding baryons. 

Reducing the effects of the SCSB to ascribing quarks a dynamical mass of about $350\,{\rm MeV}$
and verbally adding that pions are light, is, however, insufficient. In fact it is inconsistent 
to stop here: one cannot say that quarks get a constituent mass but throw out their strong 
interaction with the pion field. Constituent quarks necessarily have to interact with pions, 
as a consequence of chiral symmetry, and actually very strongly. I have had an opportunity 
to talk about it recently~\cite{D04} and shall not repeat it here. 
 
Inside baryons, quarks experience various kinds of interactions: color Coulomb, color
spin-spin (or hyperfine) and the interaction with the chiral field mentioned above.
There is also a residual contact 6-quark interaction due to instantons, which is left
after its leading piece is bosonized and goes into the chiral interactions.
It is important to know which interaction is stronger and which one is weaker
and can be disregarded in the first approximation. A simple estimate using the running 
$\alpha_s$ at typical interquark separations shows that the chiral force is, numerically, 
the strongest one. There is also a theoretical argument in its favor. Taking, theoretically,
the large-$N_c$ (the number of colors) limit has been always considered as a helpful
guideline in hadron physics. It is supposed that if some observable is stable in this academic 
limit, then in the real world with $N_c=3$ it does not differ strongly from its limiting value
at $N_c\to\infty$. There are many calculations, both analytical and on the lattice, supporting
this view. Therefore, if a quantity is known to be stable in the large-$N_c$ limit, one has to be able
to get it from physics that survives at large $N_c$. At arbitrary $N_c$, baryons are made 
of $N_c$ constituent quarks sharing the same orbital but antisymmetrized in color. 
Baryons' masses grow linearly with $N_c$ but their sizes are stable in $N_c$~\cite{Witten}. 
It means that one has to be able to obtain the quark wave function in the large-$N_c$ limit, 
and that presumably it will not differ more than by a few percent from the true wave function 
at $N_c\!=\!3$. 

When the number of participants is large, one usually applies the mean field approximation
to bound states, the examples being the Thomas--Fermi approximation to atoms and the shell 
model for nuclei. In these two examples the large number of participants are distributed 
in many orbitals or shells, whereas in the nucleon all participants are in one orbital. 
This difference is in favor of the nucleon as one expects smaller corrections from the 
fluctuations about the mean field in this case. 
%[Indeed, corrections to the Thomas--Fermi approximation are known to die out as 
%$Z^{-\frac{1}{3}}$ whereas for nucleons they die out faster as $1/N_c$.]
   
If the mean field is the color one, it has to point out in some direction in the color 
space. Hence the gluon field cannot serve as the mean field without breaking color symmetry.
The mean field can be only a color-neutral one, leaving us with the meson field as the
only candidate for the mean field in baryons. Given that the interaction of constituent
quarks with the chiral field is very strong, one can hope that the baryons' properties
obtained in the mean field approximation will not be too far away from reality. It does not
say that color Coulomb or color hyperfine interactions are altogether absent but that they
can be treated as a perturbation, once the nucleon skeleton is built from the mean chiral
field. Historically, this model of baryons~\cite{DP-CQSM} has been named the Chiral Quark 
Soliton Model, where the word ``soliton'' just stands for the self-consistent chiral field 
in the nucleon. Probably a more adequate title would be the Relativistic Mean Field Approximation 
to baryons. It should be stressed that this approximation supports full relativistic invariance 
and all symmetries following from QCD.    

%%%%%%%%%%%%%%%%
%% FIGURE 6,7 %%
%%%%%%%%%%%%%%%%
\begin{figure*}[htb]
\begin{minipage}[t]{.49\textwidth}
\resizebox{1.0\textwidth}{!}{%
  \includegraphics{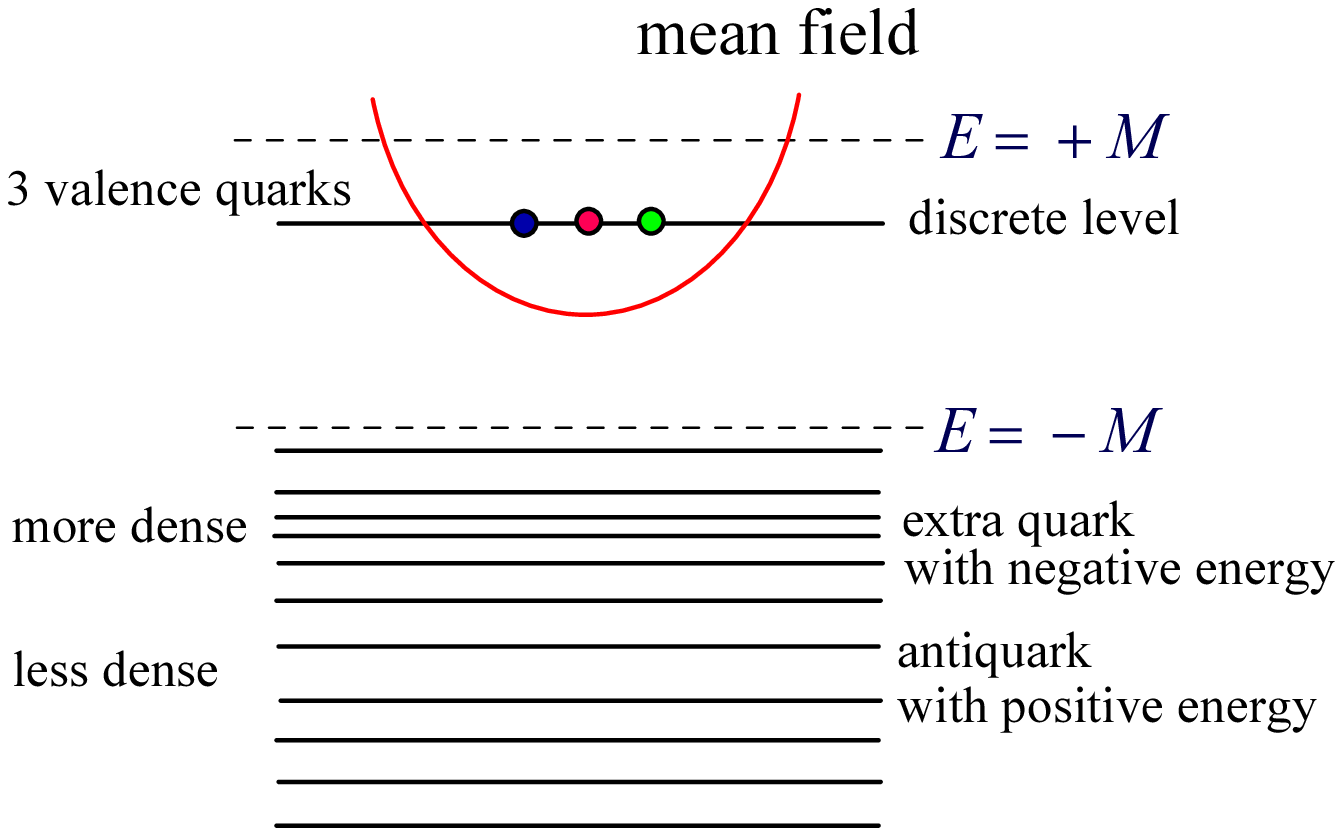}
}
\caption{A schematic view of baryons in the Mean Field Approximation. 
There are three `valence' quarks at a discrete energy level created 
by the mean field, and the negative-energy Dirac continuum distorted 
by the mean field, as compared to the free one.}
\label{fig:6}
\end{minipage}
\hspace{0.5cm}
\begin{minipage}[t]{.45\textwidth}
\resizebox{0.95\textwidth}{!}{%
  \includegraphics{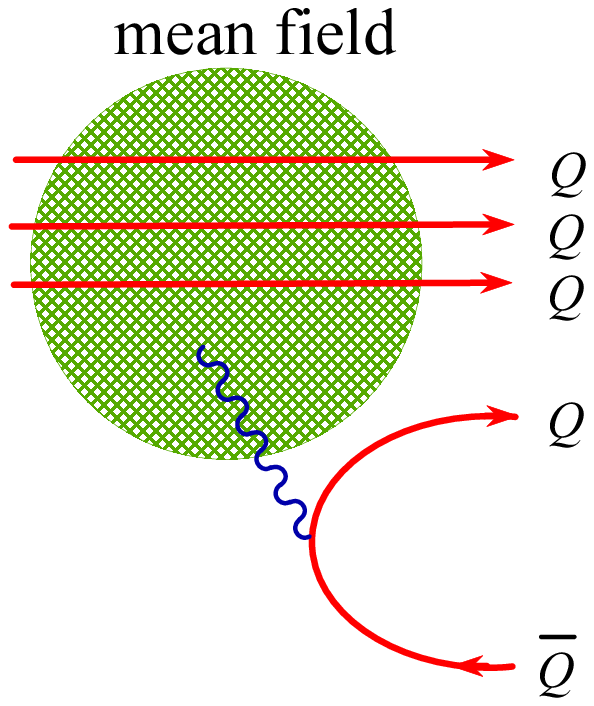}
}
\caption{Equivalent view of baryons in the same approximation, where the distorted 
Dirac sea is presented as quark-antiquark pairs. The number of $Q\bar Q$ pairs
is proportional to the square of the mean field.} 
\label{fig:7}
\end{minipage}
\end{figure*}

If the trial pion field in the nucleon is large enough (shown schematically by
the solid curve in Fig.~6), there is a discrete bound-state level for 
three `valence' quarks, $E_{\rm val}$. One has also to fill in the negative-energy
Dirac sea of quarks (in the absence of the trial pion field it corresponds
to the vacuum). The continuous spectrum of the negative-energy levels is
shifted in the trial pion field, its aggregate energy, as compared to the
free case, being $E_{\rm sea}$. The nucleon mass is the sum
of the `valence' and `sea' energies, multiplied by three colors, 
\beq
M_N=3\left(E_{\rm val}[\pi(x)]+E_{\rm sea}[\pi(x)]\right).
\la{mass}\eeq
The self-consistent mean pion field binding quarks is the one minimizing 
the nucleon mass. If it happens to be weak, the valence-quark level is shallow 
and hence the three valence quarks are non-relativistic and occupy an $s$-wave orbital. 
In this limit the Mean Field Approximation reproduces the old non-relativistic $SU(6)$ wave
functions of the octet and decuplet baryons, and there are few antiquarks~\cite{Fock}. 
If the self-consistent field happens to be large and broad, the bound-state 
level with valence quarks is so deep that it joins the Dirac sea. In this limit
the Mean Field Approximation becomes very close to the Skyrme model which
should be understood as the approximate non-linear equation for the self-consistent
chiral field. Interesting, the famous Wess--Zumino--Witten term which is added
``by hands'' in the Skyrme model~\cite{Witten-Skyrme} appears automatically~\cite{DP-CQSM}.  

The truth is in between these two limiting cases. The self-consistent pion
field in the nucleon turns out to be strong enough to produce a deep
relativistic bound state for valence quarks and a non-negligible number of antiquarks,
so that the departure from the non-relativistic quarks is considerable. 
Valence quarks are not strictly in the $s$-wave orbital but have a sizable admixture 
of the $p$-wave. At the same time the mean field is spatially not broad enough 
to justify the use of the Skyrme model which is just a crude approximation to the 
reality, although shares with reality some qualitative features. 

Being relativistic-invariant, this approach allows to compute all quark
(and antiquark) distributions in the nucleon at low virtuality where they are
not accessible in perturbative QCD. Important, all parton distributions are
positive-definite and automatically satisfy all known sum rules~\cite{SF}. 
This is because the account of the Dirac sea of quarks makes the basis states 
complete. The Relativistic Mean Field Approximation has no difficulties in explaining
the ``spin crisis''~\cite{WY} and the huge experimental value of the so-called 
nucleon $\sigma$-term~\cite{DPPrasz} -- the two stumbling blocks of the naive
quark models. Nucleon spin is carried mainly not by valence quarks but by
the orbital moment between valence and sea quarks, and inside the sea. The
$\sigma$-term is experimentally 4 times (!) bigger than it follows from
valence quarks~\cite{D04} because, again, the main contribution arises
from the Dirac sea to which the $\sigma$-term is particularly sensitive. 
On the whole, the picture of the nucleon emerging from the simple \Eq{mass}
is coherent and so far has been adequate.

\section{Nucleons under a microscope with increasing resolution}

Inelastic scattering of electrons off nucleons is a microscope with which we look into
its interior. The higher the momentum transfer $Q$, the better is the resolution of this 
microscope, see Fig.~8. 

At $q<300\,{\rm MeV}$ one does not actually discern the internal structure; it is the domain 
of nuclear physics. At $300<q<1000\,{\rm MeV}$ we see three constituent quarks 
inside the nucleon, but also additional quark-antiquark pairs; mathematically, 
they come out from the distortion of the Dirac sea in Figs.~6,7. The appropriate quark 
and antiquark distributions have been found in Ref.~\cite{SF}.
In addition, the non-perturbative gluon distribution appears for the first time
at this resolution. First and foremost, it is the glue in the interior of the constituent
quarks that has been responsible for rendering them the mass, {\it i.e.} the glue from the instanton
fluctuations. Interesting, these non-perturbative gluons are emitted not
by the vector (chromoelectric) quark current but rather by the quarks' large chromomagnetic
moment, and their distribution has been found by Maxim Polyakov and myself 
to be given by a universal function $(1-x)/x$, see section 7 in Ref.~\cite{inst-at-work}.

%%%%%%%%%%%%%%%%
%% FIGURE 8   %%
%%%%%%%%%%%%%%%%
\begin{figure}
\resizebox{0.45\textwidth}{!}{\includegraphics{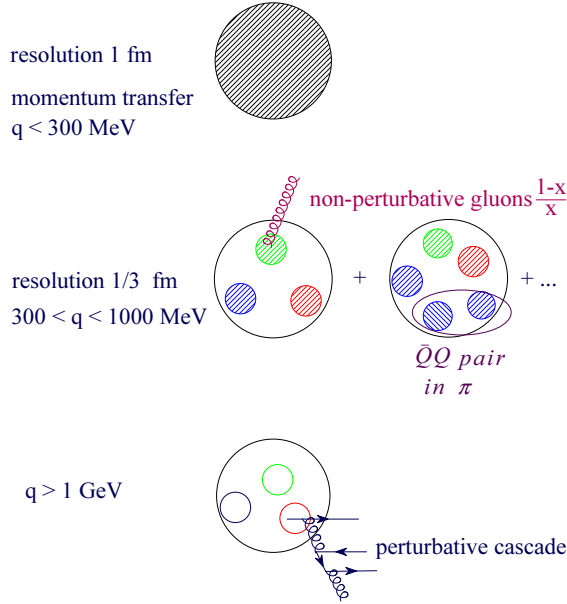}}
% If not, use
%\vspace{5cm}       % Give the correct figure height in cm
\caption{Probing the nucleon with an increasing momentum transfer $q$.}
\label{fig:8}       % Give a unique label
\end{figure}

At large $q>1\,{\rm GeV}$ one gets deep inside constituent quarks and starts to see
normal perturbative gluons and more quark-antiquark pairs arising from bremsstrahlung. 
This part of the story is well-known: the perturbative evolution of the parton cascade
gives rise to a small violation of the Bjorken scaling as one goes from moderate 
to very large momentum transfers $q$, but the basic shape of parton distributions 
serving as the initial condition for perturbative evolution, is determined at moderate 
$q$ by the non-perturbative physics described above.

\section{Pentaquarks}

Based on this picture, Victor Petrov, Maxim Polyakov and I predicted in 1997
a relatively light and narrow antidecuplet of exotic baryons~\cite{DPP97}; 
this prediction largely motivated the first experiments. Both circumstances 
-- lightness and narrowness -- are puzzles for the naive quark models. 

After the first announcements of the observation of the exotic $\Theta^+$
signal in the $\gamma\, ^{12}C$~\cite{Nakano}, $K^+{\rm Xe}$~\cite{ITEP},
$\gamma d$~\cite{CLASd} and $\gamma p$~\cite{CLASp} reactions, several theoretical
proposals appeared on how to understand pentaquarks from a traditional
constituent-quarks-only viewpoint~\cite{nrm}. 

There are basically two constituent quark approaches to pentaquarks: one of them emphasizes
the string confinement and color hyperfine interactions, the other, which I shall
call the Glozman--Riska (GR) model~\cite{GR}, stresses the pseudoscalar exchanges
as the main constituent quark interaction. Both approaches claim certain
successes in explaining the properties of the ground-state baryons and of their 
excitation spectrum. It is interesting that in order to achieve it in the GR model 
one needs to reduce the string tension by a factor of 5~(!) as compared
to that given by the lattice simulations in the pure glue world, which I find very 
natural -- see the beginning. 

I think that the truth may lie in a combination of these approaches. The moment we
say that quarks are ``constituent'', implying they have a mass of about $350\,{\rm MeV}$,
we have to acknowledge that such quarks interact strongly with the chiral field.
In this sense the GR model is right. The form of the interaction, however,
is not a simple pseudoscalar meson exchange but a fully non-linear interaction,
including 2,3... pion vertices. On top of it, the constituent quarks carry
color, hence they experience color Coulomb and color hyperfine interactions: these
may not be numerically dominant over the chiral forces but still may add to the
binding~\cite{F3}. Quarks are confined because the dynamically-generated mass $M(p)$
does not have the on-mass-shell solution $p^2+M(p^2)=0$ (as it happens in the
Schwinger model). Implementing all this in mathematical formulae is a difficult 
but not a hopeless task. In any case, I think that the Relativistic Mean Field 
Approximation is a very good starting point because it is relativistic and contains 
antiquarks as well, preserves all general symmetries and numerically seems to work well.   

Returning to pentaquarks, if one has a quark model at hand with the parameters 
fitted in the normal baryon sector, one can try to apply it to pentaquarks. 
This has started to be done, and the results are, to my mind, remarkable. 

One of the calculations is by Fl.~Stancu~\cite{nrm} in the GR model.     
Having assumed a natural color-flavor-spin-space symmetry of the pentaquark,
she has found the best variational wave function using the model parameters fixed from 
the $3Q$ baryons, and obtained the $\Theta^+$ mass. It turns out to be $510\,{\rm MeV}$
heavier than 1540 MeV. 

An evaluation of the $\Theta^+$ mass following the Jaffe--Wilczek suggestion 
of extreme diquark correlations has been recently carried out in Ref.~\cite{Simonov} 
assuming string dynamics between quarks, also fixed from fitting the usual baryons. 
The authors get $\Theta$'s mass about $0.5\,{\rm GeV}$ heavier than needed, too, 
if one assumes massless diquarks and even heavier if diquarks are not exactly 
massless~\cite{F4}. 

It is easy to understand this typical half-a-GeV overestimate of the $\Theta^+$ mass
in the constituent quark models. One sums up five quark masses each about $350\,{\rm MeV}$,
adds 150 MeV for strangeness and gets something around 1900 MeV. In addition there is
some penalty for the p-wave, assuming the $\Theta$ has positive parity. It gives more than 
2 GeV. This is the starting point. Then one switches in his or her favorite interaction 
between quarks which may reduce the starting mass, but has to pay back the kinetic energy. 
Owing to the uncertainty principle, these two usually cancel each other to a great extent, 
even if the binding force is strong. Therefore, the $\Theta^+$ mass of about 2 GeV is 
a natural and expected result in any constituent quark calculation.   

The fundamental difference with our approach to pentaquarks is seen from Fig.~6,7.
The fourth quark in the $\Theta^+$ is a higher density state in the Dirac sea: it has
a {\em negative energy} $E=-\sqrt{M^2+{\bf p}^2}$. One does not sum five quark masses
but rather $(3M+M-M)=3M$ to start with. This is because the extra $Q\bar Q$ pair in the
pentaquark is added not in the form of, say, a vector meson where one indeed adds $2M$
but in the form of a pseudoscalar Goldstone meson, which costs nearly zero energy. 
The energy penalty for making a pentaquark is exactly zero in the chiral 
limit, had the baryon been infinitely large. Both assumptions are wrong but it gives 
the idea why one has to expect light pentaquarks. In reality, to make the $\Theta^+$ from 
the nucleon, one has to create a quasi-Goldstone K-meson and to confine it inside the baryon 
of the size $\geq 1/M$. It costs roughly
\beq
m(\Theta)-m(N)\approx \sqrt{m_K^2+{\bf p}^2}\leq \sqrt{495^2+350^2}=606\,{\rm MeV}.
\la{cost}\eeq 
Therefore, one should expect the lightest exotic pentaquark around 1546 MeV. In fact one
also adds an indefinite number of light pions to cook up the $\Theta^+$. In the Dirac
language of section 3, the naive quark models attempt to make a pentaquark by adding a
particle-hole excitation or a vector meson to the nucleon whereas in the world with the 
spontaneous chiral symmetry breaking there is a cheaper possibility: to add a collective
excitation of the vacuum, {\it i.e.} the pseudoscalar meson(s). 

Some analogy can be found in the $0^{++}$ mesons. There is definitely a large $4Q$
component, say, in the $a_0(980)$. Naively, that would imply a $4M=1400\,{\rm MeV}$ mass
but $a_0$ is 400 MeV lighter, actually close to $2m_K$. This hints a resolution:
the four quarks of the $a_0$ meson are in the form of two quasi-Goldstone bosons
where all four $M$'s are eaten up. 

$\Theta^+$ is not a bound state of five good old constituent quarks: such bound states, if they
exist, necessarily have a mass about 2 GeV. At the same time it is not a $KN$ molecule -- 
first, because its size is only about $\sqrt{2}$ larger than that of the nucleon~\cite{Polyakov}, 
second, because it is an excitation of the pion field as well, third, because its coupling to 
the $KN$ state is very weak. It is a new kind of a state. What is the giant resonance or a 
rotational state of a nucleus made of? If one wants a bound-state description of the $\Theta$ 
at all cost, the closest concept I can think of is a superposition of $KN,\,K\pi N,\,K\pi\pi N...$ 
(including the scalar $\kappa N$) bound states~\cite{F5}. However, it is simpler to think 
of the $\Theta^+$ as of a rotational excitation of the mean chiral field in the 
nucleon~\cite{DPP97}. It does not mean that one needs to abandon the quark language altogether. 
On the contrary, the $\Theta^+$ has a definite $5Q$ wave function~\cite{Fock}. 
What is important, is that one of the $uudd$ quarks has a negative energy, and it is at 
the base of our low-mass prediction.

\section{Production of pentaquarks}

At present a dozen experiments have seen the $\Theta^+$ and several, 
all at high energies, have not~\cite{F6}. Apart from very different reactions, kinematical 
ranges, experimental cuts and techniques used in various experiments, which have 
to be carefully analyzed on the case-to-case basis, there might be some general 
physics behind the production (or non-produ\-ction) of exotic baryons.

We know that $\Theta$ must have an extremely small width $\sim\!1\,{\rm MeV}$ and hence a 
very small axial transition coupling~\cite{D04}. By the same argument, its couplings
to other $Q\bar Q$ currents are also unusually small, as is the magnetic
transition moment~\cite{Polyakov}. The general reason for this suppression can be understood
in the light cone quantization where only the transition to the $5Q$ component of the
nucleon is allowed, which is suppressed by itself. However, the suppression needs not
be so strong for the $\Theta^+\to N$ transition via a non-local $Q\bar QQ\bar Q$ current,
for example in the form of the scalar $K\pi$ resonance $\kappa(800)$ or just of the continuum
$K\pi,\,K\pi\pi...$ states in the GeV region, see Fig.~9.   

%%%%%%%%%%%%%%%%%%%
%% FIGURE 9,10   %%
%%%%%%%%%%%%%%%%%%%
\begin{figure*}[htb]
\begin{minipage}[t]{.47\textwidth}
\resizebox{0.99\textwidth}{!}{\includegraphics{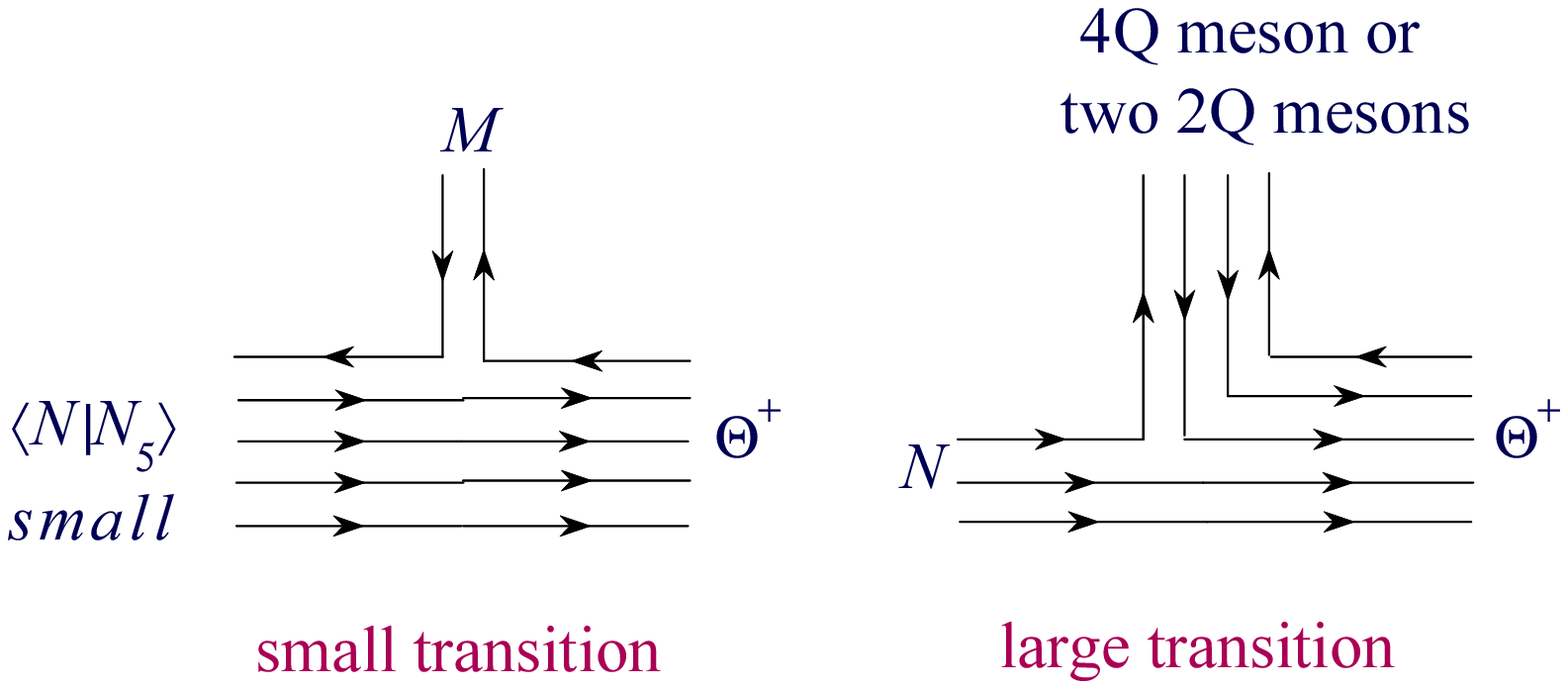}}
\caption{All transitions between a $3Q$ and a $5Q$ baryon are suppressed if mediated
by $Q\bar Q$ mesons but may not be suppressed if it is a meson with a 
large $Q\bar QQ\bar Q$ component.}
\label{fig:9}
\end{minipage}
\hspace{0.5cm}
\begin{minipage}[t]{.47\textwidth}
\resizebox{0.99\textwidth}{!}{\includegraphics{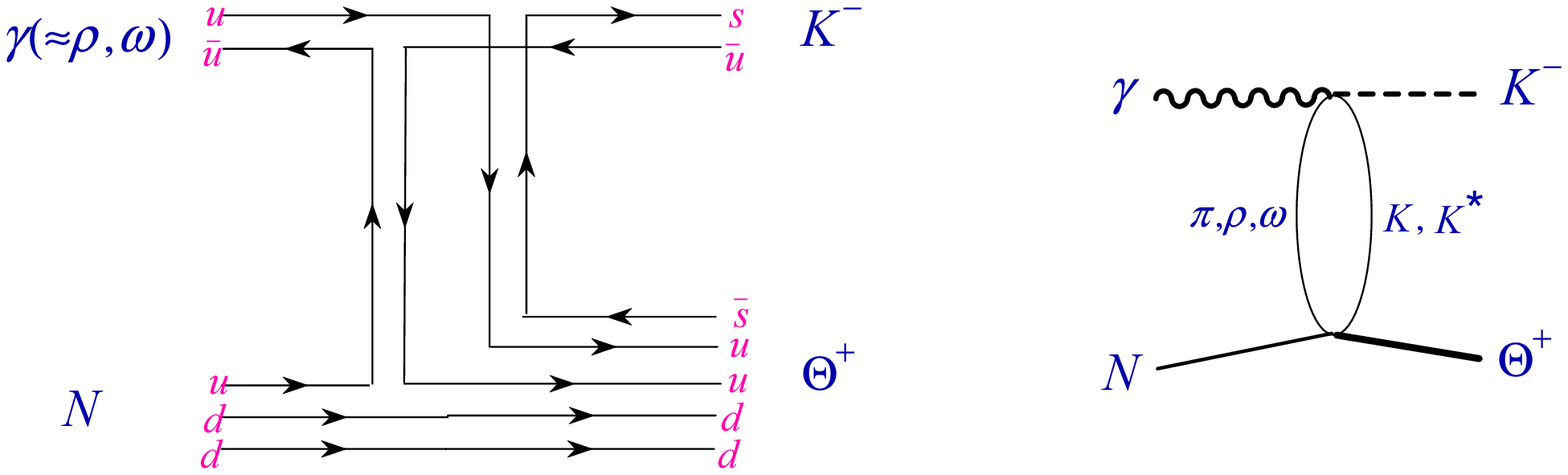}}
\caption{Possible production mechanism of the $\Theta$ at low energies. 
There is no suppression for a photon going into three mesons, for example.} 
\label{fig:10}
\end{minipage}
\end{figure*}

The $\Theta$ production mechanism at low energies could, then, look as shown in Fig.~10,
where the exchange is either of a meson with a significant $4Q$ component (like $\kappa(800)$)
or of more than one ``normal'' mesons. 

%%%%%%%%%%%%%%%%%%%
%% FIGURE 11     %%
%%%%%%%%%%%%%%%%%%%
\begin{figure*}[b]
\centerline{
\resizebox{0.60\textwidth}{!}
%\vspace*{2cm}  % Give the correct figure height in cm
{\includegraphics{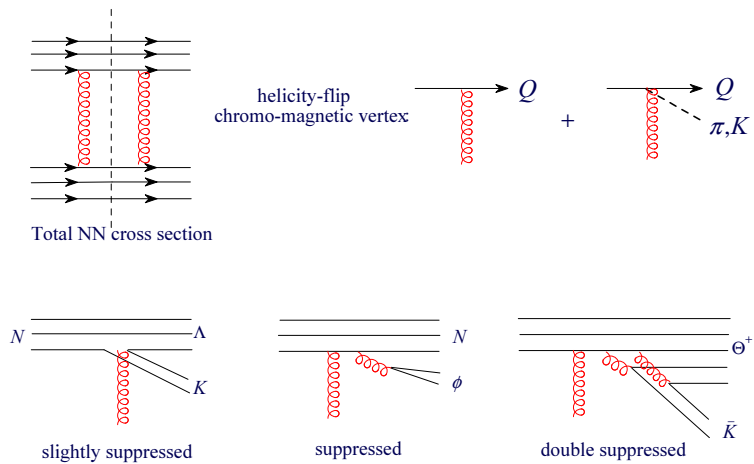}} 
}     
\caption{High energy $\Theta^+$ production from nucleon fragmentation.}
\label{fig:11}       % Give a unique label
\end{figure*} 

At high energies, all single- and double-meson exchanges die out, and only
the flavor-neutral gluon exchange survives. At low momenta transfer the gluon probably couples
to the quark via a helicity-flip chromomagnetic vertex~\cite{inst-at-work}, see Fig.~11.
In the same vertex a pion or a kaon can be easily emitted, as it
is a chirality-odd vertex. Therefore, $\Lambda K$ production at high energies is only 
combinatorially suppressed as compared to $N\pi,\,N^*\pi$ production. 
The production of the OZI-forbidden $\phi$-mesons is suppressed by about 
an order of magnitude with respect to that of $\Lambda K$, as one has to create an extra 
$Q\bar Q$ pair. $\Theta$ requires a production of two $Q\bar Q$ pairs. According to Fig.~11, 
the production of the $\Theta$ could be, roughly, in the same proportion to $\phi$ as 
$\phi$ is to $\Lambda$, or even less if one takes into account the probability of
assembling 5 quarks into the $\Theta$ wave function. Therefore, the upper limit of the 
$\Theta$ to $\Lambda(1520)$ production ratio of $10^{-2}$ found by a careful analysis 
of the SPHINX data~\cite{SPHINX} may not be altogether unexpected. Similar conclusions 
have been reached in Ref.~\cite{Titov}. To my understanding, other non-sighting experiments 
report less stringent bounds on the $\Theta$-to-$\Lambda$ production ratio.  \\

%\vskip -0.5true cm
\noindent{\bf Acknowledgments}\\
%\vspace{-0.3cm}

I thank the organizers of the Elba workshop on electron-nucleus scattering and of
the Pentaquark-04 meeting at SPring-8 for kind hospitality. I appreciate useful
discussions with S.~Brodsky, C.~Carlson, J.~Dudek, K.~Hicks, A.~Hosaka, 
H.~Lipkin, M.~Praszalowicz, S.~Stepanyan and many other people involved 
in the pentaquark epopee. This work has been supported in part by the US Department 
of Energy under contract DE-AC05-84ER40150.


\begin{thebibliography}{99}

\bibitem{F1}
Based on the talks at the workshops {\it Electron--Nucleus Scattering} (Elba, June 20-25, 2004)
and {\it Pentaquark-04} (Osaka, July 20-23, 2004), to be published in the corresponding
proceedings.

\bibitem{Schwinger}
A.~Casher, J.~Kogut and L.~Susskind, {\it Phys.~Rev.} {\bf D9}, 232 (1973); \\
G.S.~Danilov, I.T.~Dyatlov and V.Yu.~Petrov, {\it Nucl.~Phys.} {\bf B174}, 68 (1980).  

\bibitem{lin-pot}
D.~Diakonov and V. Petrov, {\it Phys.~Scripta} {\bf 61}, 536 (2000).

\bibitem{string-break}
F. Karsch, E. Laermann and A. Peikert, {\it Nucl.~Phys.} {\bf B605}, 579 (2001), {\tt hep-lat/0012023};\\
C.W.~Bernard et al., Phys. Rev. D {\bf 64}, 054506, 074509 (2001), {\tt hep-lat/0104002}; \\
A.~Duncan, E.~Eichten and J.~Yoo, {\it Phys.~Rev.} {\bf D68}, 054505 (2003);\\
H.D.~Trottier and K.Y.~Wong, {\tt hep-lat/0408028}.

\bibitem{DP-SCSB}
D.~Diakonov and V.~Petrov, {\it Phys. Lett.} {\bf B147}, 351 (1984);
{\it Nucl. Phys.} {\bf B272}, 457 (1986);

\bibitem{inst-at-work}
For a recent review on ``instantons at work'' see D.~Diakonov, {\it Prog.~Part.~Nucl.~Phys.} 
{\bf 51} (2003) 173, {\tt hep-ph/0212026}. 

\bibitem{tHooft}
G.~'t Hooft, {\it Phys. Rev. Lett.}  {\bf 37}, 8 (1976).

\bibitem{Mlat} 
P.~Bowman, U.~Heller, D.~Leinweber, A.~Williams and J.~Zhang, {\it Nucl.~Phys.~Proc.~Suppl.}
{\bf 128}, 23 (2004), {\tt hep-lat/0403002}. 

\bibitem{F2}
One may wonder if the general Dirac theory is applicable for confined quarks. 
Of course, it is: quarks in the sea are not free but interacting. Mathematically, 
one can decompose any state in plane waves or any other complete basis. 
An example of the exact description of confined electrons in the 
Schwinger model in terms of the Dirac sea is given in the second paper under 
Ref.~\cite{Schwinger}. A more fresh and relevant example is provided in:
W.~Broniowski, B.~Golli and G.~Ripka, {\it Nucl. Phys.} {\bf A703}, 667--701 (2002), 
{\tt hep-ph/0107139}.

\bibitem{Achasov}
N.N.~Achasov, {\tt hep-ph/0410051}.

\bibitem{Schaefer}
T.~Sch\"afer, {\it Phys. Rev.} {\bf D68}, 114017 (2003), {\tt hep-ph/0309158}. 

\bibitem{D04}
D.~Diakonov, talk at the APS meting (Denver, May 1-5, 2004), {\tt hep-ph/0406043}; 
talk at the {\it Continuous Advances in QCD-2004} (Minneapolis, May 12-16, 2004), 
{\tt hep-ph/0408219}, to be published in the proceedings. 


\bibitem{DP-CQSM}
D.~Diakonov and V.~Petrov, {\it JETP Lett.} {\bf 43}, 75 (1986) 
[{\it Pisma Zh. Eksp. Teor. Fiz.} {\bf 43}, 57 (1986)]; 
D.~Diakonov, V.~Petrov and P.V.~Pobylitsa, {\it Nucl. Phys.} {\bf B306}, 809 (1988);
D.~Diakonov and V.~Petrov, in {\it Handbook of QCD}, M.~Shifman, ed., 
World Scientific, Singapore (2001), vol. 1, p. 359, {\tt hep-ph/0009006}. 

\bibitem{Witten}
E.~Witten, {\it Nucl. Phys.} {\bf B156}, 269 (1979).
 
\bibitem{Fock}
D.~Diakonov and V.~Petrov, to be published in {\it Annalen der Physik}, {\tt hep-ph/0409362}.

\bibitem{Witten-Skyrme}
E.~Witten, {\it Nucl. Phys.} {\bf B160}, 433 (1983).

\bibitem{SF}
D.~Diakonov, V.~Petrov, P.~Pobylitsa, M.~Polyakov and C.~Weiss,
{\it Nucl. Phys.} {\bf B480}, 341 (1996), {\tt hep-ph/9606314}; 
{\it Phys. Rev.} {\bf D56}, 4069 (1997), {\tt hep-ph/9703420}.

\bibitem{WY}
M.~Wakamatsu and H.~Yoshiki, {\it Nucl. Phys.} {\bf A524}, 561 (1991).

\bibitem{DPPrasz}
D.~Diakonov, V.~Petrov and M.~Praszalowicz, {\it Nucl. Phys.} {\bf B323}, 53 (1989). 

\bibitem{DPP97}
D.~Diakonov, V.~Petrov and M.~Polyakov, {\it Z. Phys.} {\bf A359}, 305 (1997), 
{\tt hep-ph/9703373}; {\tt hep-ph/0404212}.  

\bibitem{Nakano}
T. Nakano (LEPS Collaboration), Talk at the PANIC 2002 (Oct. 3, 2002, Osaka);
T. Nakano et al., {\it Phys. Rev. Lett.} {\bf 91}, 012002 (2003), {\tt hep-ex/0301020}.

\bibitem{ITEP}
V.A. Shebanov (DIANA Collaboration), Talk at the Session of the Nuclear Physics 
Division of the Russian Academy of Sciences (Dec. 3, 2002, Moscow);
V.V. Barmin, A.G. Dolgolenko et al., {\it Phys. Atom. Nucl.} {\bf 66}, 1715 (2003)
[{\it Yad. Fiz.} {\bf 66}, 1763 (2003)], {\tt hep-ex/0304040}. 

\bibitem{CLASd}
S. Stepanyan, K. Hicks et al. (CLAS Collaboration), {\it Phys. Rev. Lett.} {\bf 91}, 252001 (2003),
{\tt hep-ex/0307018}. 

\bibitem{CLASp}
V. Kubarovsky et al. (CLAS Collaboration), {\it Phys. Rev. Lett.} {\bf 92}, 032001 (2004), 
{\tt hep-ex/0311046}. 

\bibitem{nrm}
Fl.~Stancu and D.-O.~Riska, {\it Phys. Lett.} \textbf{B575}, 242 (2003), {\tt hep-ph/0307010};\\ 
Fl.~Stancu, {\it Phys. Lett.} \textbf{B595}, 269 (2004), {\tt hep-ph/0402044};\\
M.~Karliner and H.~Lipkin, {\it Phys. Lett.} \textbf{B575}, 249 (2003), {\tt hep-ph/0402260};\\
R.L.~Jaffe and F.~Wilczek, {\it Phys. Rev. Lett.} \textbf{91}, 232003 (2003), {\tt hep-ph/0307341};\\
L.~Glozman, {\it Phys. Lett.} \textbf{B575}, 18 (2003), {\tt hep-ph/0308232};\\
B.~Jennings and K.~Maltman, {\it Phys. Rev.} \textbf{D69}, 094020 (2004), {\tt hep-ph/0308286};\\
R.~Bijker, M.M.~Giannini and E.~Santopinto, {\it Eur. Phys. J.} \textbf{A22}, 319 (2004), {\tt hep-ph/0310281};\\
C.E.~Carlson, C.D.~Carone, H.J.~Kwee and V.~Nazaryan, {\it Phys. Rev.} \textbf{D70}, 037501 (2004), 
{\tt hep-ph/0312325}. 

\bibitem{GR}
L.~Glozman and D.-O.~Riska, {\it Phys. Rep.} {\bf 268}, 263 (1996).

\bibitem{F3} An estimate of the Hartree and Fock contributions of color Coulomb
quark interactions to the nucleon mass shows that it lowers the mass as compared
to that obtained from the mean chiral field calculation, by a couple of hundred MeV
[D.~Diakonov and V.~Petrov (1986) (unpublished); D.~Diakonov, J.~Jaenicke and M.~Polyakov, 
preprint LNPI (1991), (unpublilshed)].

\bibitem{Simonov}
I.M.~Narodetskii, C.~Semay, B.~Silvestre-Brac and Yu.A.~Simonov, {\tt hep-ph/0409304}.

\bibitem{F4}
A direct lattice measurement of the diquark propagator has shown that its mass is larger 
than twice the constituent quark mass about 700 MeV and hence diquarks do not seem to be 
bound, see F.~Karsch et al., {\it Phys. Rev.} {\bf D58}, 111502 (1998), {\tt hep-lat/9804023}. 
It would be important to repeat this study with the current more powerful lattice techniques.

\bibitem{Polyakov}
M.~Polyakov and A.~Rathke, {\it Eur. Phys. J.} \textbf{A18} (2003) 691, {\tt hep-ph/0303138};
M.~Polyakov, in: Proceedings of Nstar-2004, and private communication.  

\bibitem{F5}
The idea of the $K\pi N$ and $\kappa(800)N$ bound states has been put forward in 
the reference below; a pilot study shows that there is a mild attraction.
P.~Bicudo and G.M.~Marques, {\it Phys. Rev.} {\bf D69}, 011503 (2004), {\tt hep-ph/0308073}; 
F.J.~Llanes-Estrada, E.~Oset and V.~Mateu, {\it Phys. Rev.} {\bf C69}, 055203 (2004), 
{\tt hep-ph/031120}; see also the contributions by Bicudo and by Oset to the {\em Pentaquark-04} 
proceedings.

\bibitem{F6}
For an experimental review see K.~Hicks, {\tt hep-ex/0412048} and the Proceedings of the Workshop
{\it Pentaquark-04}, SPring-8, Osaka, July 20-23 2004, to be published by World Scientific.

\bibitem{SPHINX}
Yu.M.~Antipov {\it et al.} (SPHINX collaboration), {\it Eur. Phys. J.} \textbf{A21} (2004) 455,
{\tt hep-ex/0407026}.

\bibitem{Titov}
A.I.~Titov, A.~Hosaka, S.~Date and Y.~Ohashi, {\tt nucl-th/0408001}. 


\end{thebibliography}
\end{document}